\documentclass[a4paper,11pt]{article}

\pdfoutput=1 

\usepackage[T1]{fontenc} 

\usepackage[]{caption}
\usepackage{color,xcolor,graphicx,amsmath, amssymb,mathtools
, physics,ulem}

\usepackage{upgreek}			
\usepackage{bm}					
\renewcommand{\ge}{g_{\star \epsilon}}			
\usepackage{booktabs} 

\newcommand{\be}{\begin{eqnarray}}
\newcommand{\ee}{\end{eqnarray}}

\newcommand{\beq}{\begin{equation}}
\newcommand{\eeq}{\end{equation}}

\definecolor{shilamagenta}{rgb}{0.8, 0.0, 0.8}

\definecolor{shilagreen}{rgb}{0.0, 0.5, 0.0}
\definecolor{shilacyan}{rgb}{0.0, 0.58, 0.71}

\definecolor{midnightblue}{rgb}{0.1, 0.1, 0.44}

\usepackage{hyperref}

\hypersetup{citecolor=blue,        
    linkcolor=red,  
    filecolor=magenta,      
    urlcolor=shilacyan  
}

\usepackage{cleveref}
\usepackage{slashed}

\begin{document}

\title{\bf {\textcolor{black}{Balancing Asymmetric Dark Matter with Baryon Asymmetry by Sphaleron Transitions}}\footnote{Presented at the 1st Electronic Conference on Universe, {22--28 February} 2021; Available online: \url{https://ecu2021.sciforum.net/}.}}

\author{Arnab~Chaudhuri$^{a}$\footnote{{\bf e-mail}: \href{mailto:arnabchaudhuri.7@gmail.com}{arnabchaudhuri.7@gmail.com}},
Maxim Yu. Khlopov$^{b}$\footnote{{\bf e-mail:} \href{mailto:khlopov@apc.in2p3.fr}{khlopov@apc.in2p3.fr}},
\\
$^a$ \small{ Novosibirsk State University} \\
\small{ Pirogova ul., 2, 630090 Novosibirsk, Russia}\\
$^b$ \small{Institute of Physics, Southern Federal University}\\
\small{Stachki 194 Rostov on Don 344090, Russia}\\
\small{and Université de Paris, CNRS, Astroparticule et Cosmologie, F-75013 Paris, France}\\
\small{and National Research Nuclear University "MEPHI", Moscow 115409, Russia}\\
}

\date{}
\maketitle

\begin{abstract}
The effect of the electroweak sphaleron transition in balance between baryon excess and and the excess of stable quarks of 4th generation is studied in this paper. Considering the non-violation of $SU(2)$ symmetry and the conservation of electroweak and new charges and quantum numbers of the new family, it makes possible sphaleron transitions between baryons, leptons and 4th family of leptons and quarks. In this paper, we have tried to established a possible definite relationship between the value and sign of the 4th family excess relative to baryon asymmetry. If $U$-type quarks are the lightest quarks of the 4th family and sphaleron transitions provide excessive $\bar U$ antiquarks,
asymmetric dark matter in the form of dark atom bound state of ($\bar{U} \bar{U} \bar{U}$) with primordial He nuclei is balanced with baryon asymmetry.
\end{abstract}

\section{Introduction}
SPHALERON 
(or electroweak sphaleron)~\cite{NSM} and \cite{KM} is a static (time-independent) solution to the electroweak field equations of the Standard Model of the particle physics. It is involved in certain hypothetical processes that violate baryon and lepton numbers. Such process cannot be represented by perturbative methods (like Feynman Diagrams) and hence termed as non perturbative.

It is possible for a sphaleron to convert baryons into antileptons and antibaryons to leptons and thus violate the baryon number. If the density of the sphaleron at some point high enough, they could wipe out any net excess of baryons or anti-baryons. Sphalerons are also associated with saddle points. It can be interpreted as the peak energy configuration. It is shown in ~\cite{3,4} that because of this anomaly transition between vacua are associated with a violation of Baryon (and lepton) numbers.

Even though a baryon excess can be created at the time of electroweak symmetry breaking, it is preserved if the process happened through first order phase transition. In second order phase transition, sphalerons can wipe out the entire asymmetry created but in first order, only the asymmetry created in the unbroken phase is wiped out.

The cosmological dark matter may contain dark atoms, in which new stable charged particles are bound by ordinary Coulomb interaction, see \cite{5,6,7,8,9}. Many models can exists in which heave $-2$ charged stable species are predicted:

1. AC-leptons, predicted in the extension of standard model, based on the approach of almost-commutative geometry, see \cite{10,11,12,13}.

2.  Technileptons and anti-technibaryons in the framework of walking technicolor models (WTC), see \cite{14,15,16,17,18,19}.

3.  and, finally, stable ”heavy quark clusters” ($\bar{U} \bar{U} \bar{U}$) formed by anti-U quark of 4th generation, see \cite{20, 21, 22}.

All these models also predict corresponding +2 charge particles. If these positively charged particles remain free in the early Universe, they can recombine with ordinary electrons in anomalous helium, which is strongly constrained in the terrestrial matter. Therefore cosmological scenario should provide a mechanism, which suppresses anomalous helium. There are two ways this can be done:

1. The abundance of anomalous helium in the Galaxy may be significant, but in the terrestrial matter there exists a recombination mechanism
suppressing this abundance below experimental upper limits.
The existence of a new strict U(1) gauge symmetry, causing new
Coulomb like long range interaction between charged dark matter
particles, is crucial for this mechanism. Therefore the existence of dark radiation in the form of hidden photons is inevitable in this approach.

2. Free positively charged particles are already suppressed in the early Universe and the abundance of anomalous helium in the Galaxy is negligible.

Dark matter production and the detailed study of sphaleron transition due to walking technicolor model has already been conducted, see \cite{18}.

In this paper, we consider 4th generation of particle family with a stable quark and neutrino family and study the sphaleron transition during electroweak phase transition. We consider the simplest model which is charge neutral and all the particles are stable. We also consider the simplest second order phase tranistion. Other complicated process such as annihilation of unstable particles and first order phase tranistion are beyond the scope of this paper.

\section{Calculations and Discussions}

The main scope of our current activity is to deduce the relationship between baryon excess and excess of stable fermions of 4th family. We can take for definiteness that 4th neutrino and U-quark of 4th family are stable and establish the relationship between their excess and excess of baryons and leptons by sphaleron transitions. The aim is to find conditions at which observed baryon excess corresponds to such an excess of $\bar{U}$ that explains observed dark matter by $(\bar{U} \bar{U} \bar{U})He$ atoms.

Three elementary particle frames are established and we are interested in the frame of a heavy quark and a heavy neutrino (neutrino with half the mass of Z boson). A Dark matter model can be proposed which can arise from this model if the baryon asymmetric universe with normal baryon also contains stable antiquarks $\bar{U}$ of 4th generation.

Owing to the $\bar{{U}}$ excess only $-2$ charge or neutral hadrons are present in the universe and $^4He$ after it is formed in Big Bang nucleosynthesis completely screens $Q^{--}$ charged hadrons in composite $[^4HeQ^{--}]$ ``atoms''. These neutral primordial nuclear interacting objects saturate the modern DM density and play an important role of the non-trivial form of the strongly interacting dark matter.

In the given scenario with electric charge $q=2/3$, the direct experimental lower limit for stable $U$-type quark is $m_U>220$ {GeV}. 
The stronger lower limit can be estimated as $m_U \ge 1$ {TeV} 
from search for R-hadrons, which have similar features as single $U$ ($\bar U$) hadrons. Due to the large chromo-Coulomb binding energy bound states for both particles and antiparticles are formed (both doublets and triplets). Even though accelerator physics excludes the formation of these bound states but they can be formed in the early universe and strongly influence the evolution of the 4th generation hadrons.

In the early universe an antibaryon asymmetry for the 4th generation of quarks can be generated so that it might form a certain fraction of dark matter. It might correspond to the modern dark matter density.

If $\epsilon(<1)$ is the fraction of dark matter then from observational data we have $\Omega_B+\epsilon (1+\Omega_{DM})\approx 0.3$ which leads to $\epsilon \approx (0.3-\Omega_B)/\Omega_{DM}$. Introducing entropy density $s$ such that $r_b=n_B/s$ and $r_{\bar{U}}=n_{\bar{U}}/s$ it is obtained that $r_b \sim 8 * 10^{-11}$ and $r_{\bar{U}}$ corresponds to the ${\bar{U}}$ excess and is a function of $\epsilon$.

A new point to note: If new stable species belong to non-trivial representation of electroweak $SU(2)$ group, sphaleron transition at high temperature can provide a relationship between baryon asymmetry and excess of `-2 charged' stable species.


If 4th family follows from string phenomenology, we have new charge
associated with 4th family fermions (which follows from rank 6 of E6
group, embedding symmetry of the Standard model with rank 4) and this
charge F should be used instead of technibaryon number TB and
technilepton number L'. In principle there should be evidently only one conserved number F for the 4th family, however, it may be easier to have analogy with WTC and assume two numbers FB (for 4th quark) and L' for (4th neutrino). Detailed calculations of WTC have been done in \cite{12,18} and most of the terminology are kept same as the above mentioned papers.

At some energy scale higher than the electroweak one, we assume the existence of a mechanism leading to a 4th asymmetry in the Universe. Given that the 4th and baryon number have a very similar nature such an asymmetry is very plausible and can have a common origin.

The asymmetry generated by 4th generation above electroweak phase transition can be destroyed by quantum anomalies. However, baryon, lepton and 4th number are not conserved even though their differences are conserved. This fact allows for a nonzero asymmetry to
survive. The processes leading to such a violation are termed “sphaleron” processes and at the present time are negligible. However these processes were active during the time the Universe had a temperature above or at the scale of the electroweak symmetry breaking.

At some points when as the Universe expands and its temperature falls the quantum number violation ceases the effect. The relation among the particles emerging from the process (SM + 4th generation):
\begin{equation} \label{5}
3(\mu_{u_L}+\mu_{d_L})+\mu+\mu_{U_L}+\mu_{D_L}+\mu_{L'}=0
\end{equation}
$\mu$ is chemical potential summed over all standard model leptons  and $\mu_{L'}$ is the new generation leptons (and neutrinos). $\mu_{U_L}$ and  $\mu_{D_L}$ are the chemical potential for the 4th generation quarks. For detailed calculations and breakdown of each parameter, see \cite{18}.
{The number density follows, for fermions and bosons, respectively, the general expression:}
\begin{equation} \label{6}
n=g_*T^3\frac{\mu}{T}f(\frac{m}{T})
\end{equation}
and
\begin{equation} \label{7}
n=g_*T^3\frac{\mu}{T}g(\frac{m}{T})
\end{equation}
where $f$ and $g$ are hyperbolic functions and $g_*$ is the effective degrees of freedom. They are given by:
\begin{equation} \label{7a}
f(z)=\frac{1}{4\pi^2}\int_0^{\infty}x^2 cosh^{-2}\left(\frac{1}{2}\sqrt{z^2+x^2} \right)dx
\end{equation}
and
\begin{equation} \label{7b}
g(z)=\frac{1}{4\pi^2}\int_0^{\infty} x^2 sinh^{-2}\left(\frac{1}{2}\sqrt{z^2+x^2} \right)dx
\end{equation}

The Baryon number density is defined as the following:
\begin{equation} \label{8}
B=\frac{n_B-n_{\bar{B}}}{gT^2/6}
\end{equation}

This normalization is used thorough out and since in the end the ratio is of interest, this normalization cancels out.

Defining the parameter $\sigma$, respectively, for fermions and bosons:
\begin{equation} \label{9}
\sigma=6f\frac{m}{T^*}
\end{equation}
and
\begin{equation} \label{10}
\sigma=6g\frac{m}{T^*}
\end{equation}
where $T^*$ is the phase transition temperature and is given by the following expression:
\begin{equation} \label{10a}
T^*=\frac{2M_W(T^*)}{\alpha_Wln(M_{pl}/T^*)}B(\frac{\lambda}{\alpha_W})
\end{equation}
where $M_W$ is the mass of W-boson, $M_{pl}$ is the Planck mass and $\lambda$ is the self-coupling. The function $B$ is derived from experiment and takes the value from 1.5 to 2.7.

The new generation charge is calculated to be:
\begin{equation} \label{11}
FB=\frac{2}{3}(\sigma_{U_L}\mu_{U_L}+\sigma_{U_L}\mu_{D_L}+\sigma_{D_L}\mu_{D_L}).
\end{equation}
where $FB$ corresponds to the anti-U ($\bar{U}$) excess.

The standard model baryonic charge and leptonic charges are given by the following expression:
\begin{equation} \label{10a1}
B=\left[(2+\sigma_t)(\mu_{uL}+\mu_{uR})+3(\mu_{dL}+\mu_{dR}) \right]
\end{equation}
and
\begin{equation} \label{10b}
L=4\mu+6\mu_W
\end{equation}
where in Equation (\ref{10a1}) 
the factor 3 of the down-type
quarks is the number of families. For the new lepton family, originating from 4th generation, we have:
\begin{equation} \label{10c}
L'=2(\sigma_{\nu'}+\sigma_{U_L})\mu_{\nu'L}+2\sigma_{U_L}\mu_W+(\sigma_{\nu'}-\sigma_{U_L})\mu_0
\end{equation}
where $\nu'$ is the new family of neutrinos originating from the extension of standard model and $\mu_0$ is the chemical potential from the Higgs sector in standard model.

In a second order phase transition we expect the violating processes to persist below the phase transition and the equilibrium conditions are imposed after the phase
transition.  If the phase diagram as function of temperature and density of the 4th generation would be known a specific order of the electroweak phase transition would be used. The ratio of the number densities of 4th generation to the baryons is determined by:
\begin{equation} \label{12}
\frac{\Omega_{FB}}{\Omega_B}=\frac{3}{2}\frac{FB}{B}\frac{m_{FB}}{m_p}
\end{equation}

For second order phase transition, the electrical neutrality and zero chemical potential of the Higgs. The ratio of the number density of number density of the 4th generation to the baryon number density can be expressed as a function of $L/B$ and $L'/B$. In the limiting case, we get:
\begin{equation} \label{13}
-\frac{FB}{B}=\frac{\sigma_{U_L}}{3(18+\sigma_{\nu'})}\left[(17+\sigma_{\nu'})+\frac{(21+\sigma_{\nu'})}{3}\frac{L}{B}+\frac{2}{3}\frac{9+5\sigma_{\nu'}}{\sigma_{\nu'}}\frac{L'}{B}\right].
\end{equation}

Hence a possible relationship between the baryon excess and the excess of $\bar{U}$ has been established in the framework of second order electroweak phase transition.




\section{Conclusions}

It is clear from Equation (\ref{13}) that a definite relationship between the value and sign of 4th generation family excess relative to the baryon asymmetry due to electroweak phase transition and possible sphaleron production has been established. Dark matter candidates in the form of bounded dark atom can emerge from this model due to the excess of $\bar{U}$ within primordial He nuclei. We have considered only the lightest and most stable particles and also took into account only the second order phase transition. The situation can change more drastically if unstable particle and first order phase transition were taken into account but those scenarios are beyond the scope of this paper.



\section{Author Contribution}
Article by A.C. and M.K. The authors contributed equally to this~work.
All authors have read and agreed to the published version of the manuscript.

\section{Funding}
The work of A.C. is funded by RSF Grant 19-42-02004. The research by M.K. was financially supported by Southern Federal University, 2020 Project VnGr/2020-03-IF. 

\section{Conflict of Interest}
There has been no conflict of interest among the authors of this paper.

\end{document}